\documentclass[reprint, superscriptaddress, floatfix, amsmath, amssymb, aps, prx, longbibliography]{revtex4-2}

\usepackage{graphicx}
\usepackage{dcolumn}
\usepackage{makecell}
\usepackage{multirow}
\usepackage{placeins}
\usepackage{float}
\usepackage{hyperref}
\usepackage{placeins}
\usepackage{times}
\usepackage{physics}
\usepackage{siunitx}
\usepackage{mathtools}
\usepackage[dvipsnames]{xcolor}
\usepackage{float}
\usepackage{lineno}
\hypersetup{
    colorlinks,
    citecolor=blue,
    filecolor=black,
    linkcolor=blue,
    urlcolor=black
}

\renewcommand{\figurename}{\textbf{Fig.}}
\renewcommand{\tablename}{\textbf{Table}}
\makeatletter
\def\fnum@figure{\figurename\nobreakspace\textbf{\thefigure}}
\def\fnum@table{\tablename\nobreakspace\textbf{\thetable}}
\makeatother

\begin{document}

\title{Quantum Cellular Automata on a Dual-Species Rydberg Processor}

\author
{\normalsize{
Ryan White,$^{1}$$^\dag$
Vikram Ramesh,$^{1}$$^\dag$
Alexander Impertro,$^{2}$
Shraddha Anand,$^{3}$
Francesco Cesa,$^{2,4}$
Giuliano Giudici,$^{2,4}$
Thomas Iadecola,$^{5,6,7}$
Hannes Pichler,$^{2,4}$
and Hannes Bernien$^{2,3,8\ast}$
}\\
\small{$^{1}$Department of Physics, University of Chicago, Chicago, IL 60637, USA}\\
\small{$^{2}$Institute for Quantum Optics and Quantum Information of the Austrian Academy of Sciences, 6020 Innsbruck, Austria}\\
\small{$^{3}$Pritzker School of Molecular Engineering, University of Chicago,  Chicago, IL 60637, USA}\\
\small{$^{4}$Institute for Theoretical Physics, University of Innsbruck, 6020 Innsbruck, Austria}\\
\small{$^{5}$Department of Physics, The Pennsylvania State University, University Park, Pennsylvania 16802, USA}\\
\small{$^{6}$Institute for Computational and Data Sciences, The Pennsylvania State University, University Park, Pennsylvania 16802, USA}\\
\small{$^{7}$Materials Research Institute, The Pennsylvania State University, University Park, Pennsylvania 16802, USA}\\
\small{$^{8}$Institute for Experimental Physics, University of Innsbruck, 6020 Innsbruck, Austria}\\
\small{$^\dag$These authors contributed equally to this work.} \\
\small{$^\ast$To whom correspondence should be addressed; E-mail:~hannes.bernien@uibk.ac.at}
}

\date{\today}

\begin{abstract}

As quantum devices scale to larger and larger sizes, a significant challenge emerges in scaling their coherent controls accordingly~\cite{preskill2018quantum, awschalom2025challenges}. Quantum cellular automata (QCAs)~\cite{farrelly2020review} constitute a promising framework that bypasses this control problem: universal dynamics can be achieved using only a static qubit array and global control operations~\cite{margolus1986quantum, lloyd1993potentially, watrous1995onedimensional, benjamin2001simple, levy2002universal, raussendorf2005qca}. Despite an extensive history of theoretical explorations and proposals, QCAs have not been experimentally explored in the context of highly-scalable globally-controlled systems. Here we realize QCAs on a dual-species Rydberg array of rubidium and cesium atoms, leveraging independent global control of each species to perform multiple quantum protocols. Seeding an automaton with different initial states, we explore many-body dynamics of quasiparticles and grow GHZ states across both species, highlighting the flexibility of our approach. We further develop a second automaton using a novel mediated entangling gate, enabling generation of 96.7(1.7)\%-fidelity Bell states, 17-qubit cluster states, and high-connectivity graph states~\cite{schlingemann2001qec}. Our results demonstrate that simple global controls enable access to a rich variety of applications through the QCA framework. The versatility and scalability of QCAs present compelling opportunities for the development of quantum information systems~\cite{cesa2023universal}, as well as new perspectives on quantum many-body dynamics~\cite{iadecola2020nonergodic, giudici2024unraveling}.

\end{abstract}

\maketitle

\section*{Introduction}

Coherently controlling large quantum systems is an outstanding challenge in the development of next-generation quantum devices~\cite{preskill2018quantum, awschalom2025challenges}. Some of the earliest quantum processing proposals foresaw this difficulty, and designed protocols which could achieve universal performance using only global controls and locally-interacting qubits~\cite{margolus1986quantum, lloyd1993potentially, watrous1995onedimensional}. With the advent of highly scalable qubit architectures, such as neutral atom arrays~\cite{saffman2010qi, bluvstein2024logical, manetsch2025tweezer}, these protocols are particularly attractive as a way to limit the control complexity.   

\begin{figure*}
\centerline{\includegraphics[scale=1]{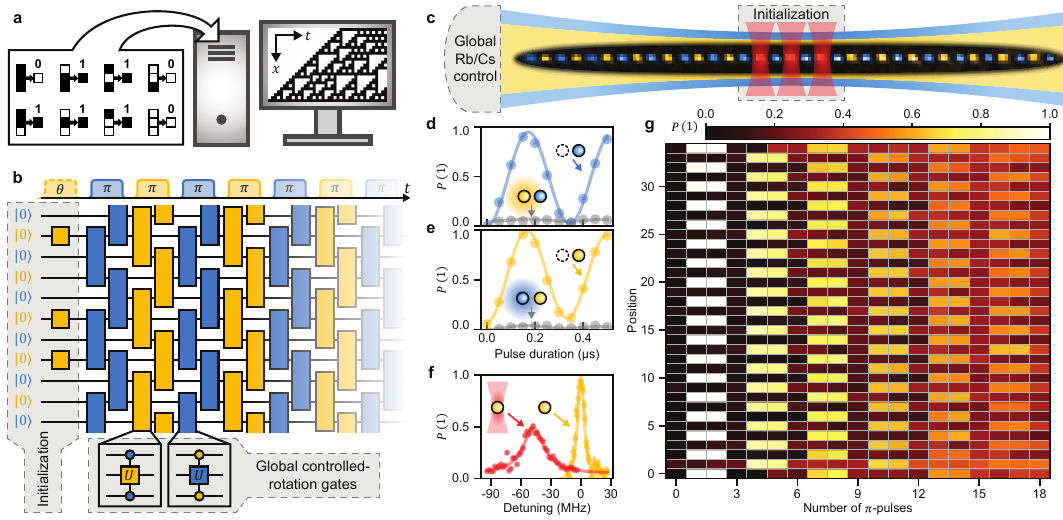}}
\caption{\textbf{Dual-species quantum cellular automata.} \textbf{a}, Classical cellular automata repeatedly apply an update rule to a grid of ``cells", leading to complex dynamics from simple initial states. \textbf{b}, Replacing the classical cells and update rules with qubits and unitary steps (e.g.~controlled qubit rotations) yields a quantum cellular automaton (QCA). These unitary steps can be applied by global controls, facilitating simple manipulation of complex quantum systems from a separable initial state. \textbf{c}, Averaged fluorescence image of our rubidium (blue) and cesium (yellow) Rydberg qubit array, which is used to implement QCAs. Unitary operations are applied via global lasers, and AOD tweezers (red) are used to control initialization. \textbf{d}, \textbf{e}, Rabi oscillations between the ground and excited states (blue/yellow data) are driven by species-selective lasers. Presence of a neighboring atom in the Rydberg state induces a blockade effect, preventing excitation of the driven atom (gray data). Solid lines are fits to damped cosines. \textbf{f}, Spectroscopy of the Rydberg transition reveals that the free-space resonance frequency (yellow data) can be strongly shifted using AOD tweezers (red data), shielding selected atoms during initialization. Solid lines are fits to Lorentzians. \textbf{g}, Applying alternating Rb/Cs $\pi$-pulses on the 35-qubit array implements the PXP automaton. Tracking the Rydberg state population over time, we observe a periodic ``vacuum orbit", where at each time step at most one species is excited.}
\label{fig:setup}
\end{figure*}

Many of these globally-controlled protocols fall within the scope of quantum cellular automata (QCAs). In \emph{classical} cellular automata, such as Wolfram’s Rule 110~\cite{wolfram2003new} (Fig.~\ref{fig:setup}a) and Conway’s Game of Life~\cite{gardner1970mathematical}, vast functionalities and complex behaviors arise from simple global controls and a properly-chosen initial state. Here, discrete evolution of a grid of ``cells" under a local update rule yields highly complex dynamics~\cite{vonneumann1966theory, chopard1998cellular} and has been shown to be Turing complete~\cite{cook2004universality}. Correspondingly, \emph{quantum} cellular automata use the repeated application of translation-invariant multi-qubit unitary operations to evolve a grid of qubits~\cite{farrelly2020review} (Fig.~\ref{fig:setup}b) and can enable universal quantum computation~\cite{margolus1986quantum, lloyd1993potentially, watrous1995onedimensional, benjamin2001simple, levy2002universal, raussendorf2005qca}. Furthermore, QCAs provide new ways to study quantum many-body dynamics, including phase transitions in critically-tuned systems, quantum many-body scars, and quantum chaos~\cite{gillman2020nonequilibrium, gillman2021correlations, iaconis2020measurement, iaconis2019anomalous, gopalakrishnan2018operator, rozon2022constructing}. Recent experiments have demonstrated the feasibility of implementing QCAs, by simulating single-particle Dirac quantum walks on trapped ion and photonic platforms~\cite{alderete2020quantum, suprano2024photonic} and investigating complex network generation on superconducting qubits~\cite{jones2022smallworld}. 

In this work, we implement quantum cellular automata on Rydberg arrays of up to 35 neutral-atom qubits, realizing the first dual-species experiment to probe interacting many-body dynamics. The inherent independent addressability of distinct atomic species~\cite{singh2023midcircuit}, combined with a nearest-neighbor Rydberg blockade~\cite{anand2024dual}, enables access to a wide range of protocols using only global controls~\cite{cesa2023universal}. With this simple architecture, we investigate quasiparticle dynamics in a PXP automaton~\cite{wilkinson2020exact, iadecola2020nonergodic, giudici2024unraveling}, and explore deviations from the integrable regime of this model. Building on these capabilities, we use QCAs as an entanglement generation tool, producing GHZ states on both species, and design a mediated-gate protocol to generate Bell states, a 17-qubit 1D cluster state, and high-connectivity graph states. These experiments highlight that a minimal set of global controls suffices to realize a plethora of advanced quantum protocols, opening new doors for globally-controlled explorations of quantum information and simulation.

\section*{Implementing quantum cellular automata}

Our QCA experiments are performed on one-dimensional dual-species arrays of up to 35 qubits, alternating between rubidium (Rb, blue) and cesium (Cs, yellow) atoms (Fig.~\ref{fig:setup}c). As in previous work~\cite{singh2023midcircuit, anand2024dual}, we use two spatial light modulators (SLMs) to independently trap single atoms of the respective element. We further implement dual-species rearrangement using a single pair of crossed acousto-optic deflectors (AODs) to create defect-free arrays~\cite{supp}. Information is encoded in ground-Rydberg qubits, where the ground states ($\ket{0}_\text{Rb} = \ket{5S_{1/2}, F=2, m_F=-2}$ and $\ket{0}_\text{Cs} = \ket{6S_{1/2}, F=4, m_F=-4}$) are prepared by optical pumping, and the Rydberg states ($\ket{1}_\text{Rb} = \ket{67S_{1/2}, m_J=-1/2}$ and $\ket{1}_\text{Cs} = \ket{67S_{1/2}, m_J=-1/2}$) are accessed with species-selective lasers~\cite{supp}. Our atomic spacing of 5.3\,\textmu m gives a strong nearest-neighbor blockade (Fig.~\ref{fig:setup}d,e), which we use to implement controlled-rotation unitaries. Finally, we use AC Stark shifts from AOD-generated tweezers to strongly detune the Rydberg state at selected sites, preventing resonant excitation (Fig.~\ref{fig:setup}f). This addressing can be used for a single pulse, acting as either an initialization step for the QCA or a way to measure a subset of atoms in a different basis.

Using these capabilities, the first QCA we demonstrate consists of a sequence of Rydberg $\pi$-pulses, alternating between Rb and Cs at each step. In isolation, these pulses would flip the state of their respective species at each time step, but because of the strong nearest-neighbor Rydberg blockade, the flip becomes controlled by the state of neighboring atoms. Ignoring higher-order effects from the van der Waals interaction, this sequence can be interpreted as a discretized simulation of the PXP model~\cite{bernien2017probing, turner2018scarred}. The unitary step for an even(odd)-numbered pulse can be written as~\cite{iadecola2020nonergodic}
\begin{equation} \label{eqn:pxp}
    U_\text{even(odd)} = \prod_{j \text{ even(odd)}}{e^{-i \frac{\pi}{2} P_{j-1} X_j P_{j+1}}},
\end{equation}
where $X_j$ is the Pauli X operator on the $j$th qubit, and $P_j = \ket{0}\bra{0}_j$ is a projector that prevents its neighbors from evolving unless qubit $j$ is in the ground state.

When we repeatedly apply this unitary step to a full 35-atom chain initialized in the ground state, $\ket{\text{vacuum}}=\ket{0000...}$, we see the following dynamics: first, all of the Rb atoms are flipped, creating the state $\ket*{\mathbb{Z}_2^\text{(even)}} = \ket{1010...}$; next, the Rb-Cs blockade prevents Cs from flipping; then, Rb is brought back to the ground state ($\ket{\text{vacuum}}$), allowing Cs to be subsequently flipped ($\ket*{\mathbb{Z}_2^\text{(odd)}} = \ket{0101...}$), and the pattern then repeats on the other species (Fig.~\ref{fig:setup}g). This short cycle, which repeats every six $\pi$-pulses, is referred to as a ``vacuum orbit". Its presence signals the existence of nonthermal eigenstates in the periodically driven quantum system~\cite{iadecola2020nonergodic}, which in turn bear a connection to scarred eigenstates of the PXP model~\cite{giudici2024unraveling}. Loss of contrast over time is influenced by several factors, including van der Waals interaction tails, calibration of system parameters, and decoherence processes~\cite{supp}.

\begin{figure*}
\centerline{\includegraphics[scale=1]{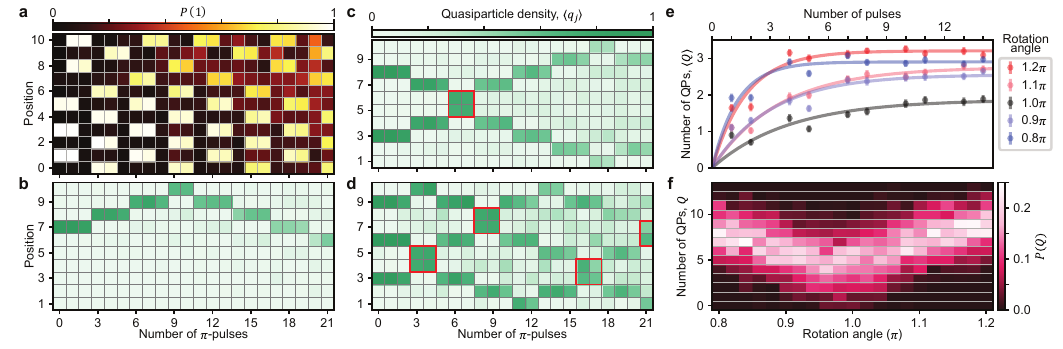}}
\caption{\textbf{Quasiparticles in the PXP QCA.} \textbf{a}, A domain wall state is initialized and then evolved under global PXP unitary steps. \textbf{b}, By identifying the position of the domain wall in each shot of the experiment, we build a quasiparticle position histogram for each time step, revealing its linear motion and reflection. Initializing states with more domain walls, we further observe the dynamics of \textbf{c}, two- and \textbf{d}, three-quasiparticle configurations. When two quasiparticles collide (red squares), there is an interaction which modifies their trajectory. \textbf{e}, By over- or under-rotating our global $\pi$-pulses, we deviate from the integrable regime of the PXP automaton. For a 15-atom chain initialized in $\ket{0}$, increasing the deviation from $\pi$ results in an increasingly fast saturation of the quasiparticle number. Exponential fits are a guide to the eye. \textbf{f}, Looking at a single time step (6 $\pi$-pulses) on a 35-atom array, we see the quasiparticle number distribution shifts further upwards as the rotation angle deviates from $\pi$. We note a small horizontal shift, which we attribute to a slight miscalibration of the pulses.}
\label{fig:quasiparticles}
\end{figure*}

\section*{Quasiparticles in the PXP automaton}

Analogously to classical cellular automata, exploration of QCA dynamics often necessitates preparing an initial state before applying unitary steps. This can be used, for example, to explore different dynamical regimes of the automaton, or to encode information for subsequent computation~\cite{farrelly2020review}. We realize this initialization using our AOD light shifts, applying a single pulse to prepare only the unshifted Cs atoms in $\ket{1}$~\cite{supp}. For the PXP QCA, domain walls between different vacuum configurations are particularly interesting; for example, if the left half of an array is in the state $\ket{\text{vacuum}}$, while the right half is $\ket*{\mathbb{Z}_2^\text{(odd)}}$, the domain wall lies at the interface of these sub-arrays. As unitary steps are applied, this domain wall will move across the array with fixed velocity, acting as a quasiparticle~\cite{iadecola2020nonergodic, wilkinson2020exact}.

We start by initializing a single domain wall state on arrays of 11 atoms. Applying the PXP unitary step for up to 21 $\pi$-pulses, we observe a dark region traced out in the population plot, which moves towards the top of the array, then reflects back (Fig.~\ref{fig:quasiparticles}a). To better visualize the quasiparticle behavior, we project measurement outcomes onto a set of 4-atom occupation bitstrings which signal the presence of a quasiparticle at a given location~\cite{supp}. Conditioning on instances where only a single quasiparticle was flagged (the QCA should ideally conserve the total quasiparticle number), we collect a histogram of the observed positions at each time step, and confirm the linear motion and reflection of the quasiparticle (Fig.~\ref{fig:quasiparticles}b).

Next, we prepare two- (Fig.~\ref{fig:quasiparticles}c) and three-particle (Fig.~\ref{fig:quasiparticles}d) states by initializing more domain walls. As before, we catalog quasiparticle positions over time, conditioned on instances where the targeted quasiparticle number is conserved. We observe that when quasiparticles collide with each other, they experience an interaction which modifies their subsequent trajectory: independent quasiparticles take three $\pi$-pulses to move a single step, whereas upon collision with another quasiparticle, only two $\pi$-pulses are taken.

\begin{figure*}
\centerline{\includegraphics[scale=1]{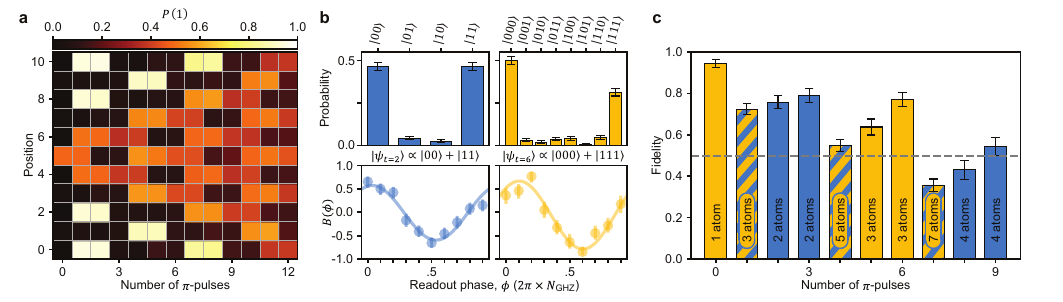}}
\caption{\textbf{Growing GHZ states with the PXP QCA.} \textbf{a}, When a single qubit is initialized in a superposition state (here, $\ket{+}=(\ket{0}+\ket{1})/\sqrt{2}$), applying PXP unitary steps causes the superposition to spread out across the array, growing a GHZ state. \textbf{b}, GHZ state fidelity is probed at each time step using population (top) and parity (bottom) measurements, shown here for time steps 2 (left, Rb) and 6 (right, Cs). Solid lines are fits to cosines. \textbf{c}, Plotting the GHZ state fidelity versus number of pulses, we verify entanglement over many time steps, up to a size of 4 single-species qubits (solid bars) or 5 dual-species qubits (striped bars).}
\label{fig:ghz}
\end{figure*}

Using only $\pi$-pulses, we remain close to the well-understood integrable regime of the PXP automaton. However, the continuous tunability of our rotation angles allows us to over- or under-rotate both species and explore the onset of non-integrable behavior. For example, the PXP automaton should conserve the total quasiparticle number, whereas over- and under-rotations would introduce quasiparticle creation and annihilation terms. We confirm this effect on arrays of 15 atoms by applying up to 14 pulses with a set rotation angle. As the angle deviates further from $\pi$, quasiparticles begin to proliferate more rapidly within the system (Fig.~\ref{fig:quasiparticles}e). To see this effect with higher resolution, we look at a fixed early time (6 $\pi$-pulses) on a full 35-atom array (Fig.~\ref{fig:quasiparticles}f). When the rotation angle is varied, we observe a clear upward shift in the distribution of quasiparticle numbers. The shift is minimized close to a rotation angle of $\pi$, corresponding to the ideal PXP automaton. Although the classical limit of the PXP automaton is well-understood, adding this rotation angle as a ``quantum knob" opens opportunities for investigations of nonergodic effects relating to the PXP vacuum orbit~\cite{iadecola2020nonergodic}.

\section*{Growing GHZ states}

We next investigate the PXP automaton as a tool for preparing entangled quantum states. Even though this QCA maps computational states to computational states, entanglement \emph{can} be generated by seeding the initial state with superposition. For instance, if we initialize one or more ``seed" qubits in the $\ket{+}=(\ket{0}+\ket{1})/\sqrt{2}$ state, applying the unitary step causes the superposition to spread out across the array, growing a GHZ state from each seed. We note that this approach to state preparation is qualitatively distinct from protocols that have been used in Rydberg array experiments previously~\cite{omran2019generation, senoo2025highfidelity}, relying on discrete pulse sequences rather than adiabatic sweeps.

Using our AOD light shifts, we prepare a single Cs atom in the $\ket{+}$ state in arrays of 11 atoms. Applying PXP unitary steps and observing the Rydberg population over time, we see a light cone form of sites with $P(1)\approx 50$\%, as expected for a GHZ state (Fig.~\ref{fig:ghz}a). Another way of interpreting this light cone is that the $\ket{0}$-component of the $\ket{+}$ state results in our standard vacuum dynamics, while the $\ket{1}$-component leads to two quasiparticles traveling away from each other. The superposition of this two-particle evolution and the vacuum evolution yields a GHZ state, located between the two quasiparticles.

We determine the fidelity of the GHZ states using standard measurements of the population and the parity $B(\phi)=\langle R^{\otimes N}(\phi)\rangle$ (Fig.~\ref{fig:ghz}b), where the readout basis $R(\phi)=\cos{(\phi)}X+\sin{(\phi)}Y$ is set with a $\pi/2$-pulse prior to readout~\cite{supp}. For most time steps, the GHZ state is supported on a single species (indicated by the color of the data in Fig.~\ref{fig:ghz}b and c) while the other species should be entirely in $\ket{0}$. Since our dual-species system enables auxiliary-qubit readout without disrupting the data qubits~\cite{singh2023midcircuit}, a straightforward strategy to increase the fidelity of the data qubits is to flag on successful return of the auxiliary qubits to $\ket{0}$. Applying this flagging technique in postselection, we confirm entanglement of the GHZ state up to a size of 4 atoms (Fig.~\ref{fig:ghz}c).

At time steps 1, 4, and 7, a dual-species GHZ state is produced. Since the species are no longer separable, flagging is not possible, and parity measurements are obstructed by the nearest-neighbor Rydberg blockade. To obtain an estimate of the fidelity at these steps, we determine a lower-bound using the parity of later time steps~\cite{supp}. This fidelity estimate (Fig.~\ref{fig:ghz}c, striped bars) indicates we are above the entanglement threshold for a 5-qubit GHZ state distributed across both Rb and Cs atoms.

\section*{Mediated gate and cluster states}

\begin{figure}
\centerline{\includegraphics[scale=1]{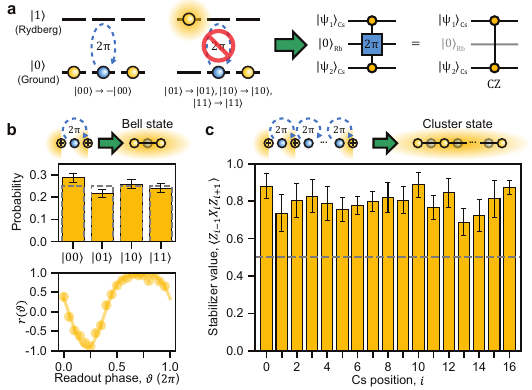}}
\caption{\textbf{Bell and Cluster states.} \textbf{a}, Applying a $2\pi$-pulse to one species (here Rb) leaves all computational states unchanged, but provides a phase depending on the state of neighboring atoms. This effects a CZ gate on the Cs atoms, and the Rb atom(s) can be traced out. \textbf{b}, With two Cs atoms in the $\ket{+}$ state, a $2\pi$ pulse on a central Rb atom creates the Bell state $(\ket{1+}-\ket{0-})/\sqrt{2}$. Population (top) and coherence (bottom) measurements indicate a Bell state fidelity of 96.7(1.7)\%. Solid line is a fit to a model function parametrized by realistic error sources~\cite{supp}. \textbf{c}, Using the same sequence on a 33-atom chain generates a 17-qubit 1D cluster state. Pauli stabilizer measurements on the Cs atoms verify entanglement across any cut of the Cs chain~\cite{toth2005entanglement}.}
\label{fig:cluster}
\end{figure}

Under strong Rydberg blockade, two adjacent atoms cannot both be in the $\ket{1}$ state, reducing the accessible Hilbert space. We circumvent this limitation by developing a mediated gate protocol, using Rb as auxiliary qubits while keeping the Cs data qubits outside of each others blockade radius, opening up the full $2^N$-dimensional Hilbert space of the data qubits. As depicted in Fig.~\ref{fig:cluster}a, with Rb starting in $\ket{0}$, a $2\pi$-pulse always returns Rb to $\ket{0}$, with either a phase of -1 in the absence of any blockade, or a phase of 1 in the presence of a strong blockade. With two neighboring Cs atoms, only their $\ket{00}$ state will pick up the $-1$ phase, realizing a CZ gate on Cs mediated by Rb. In practice, to increase qubit coherence and provide better control of pulse axes, we use a modified pulse sequence, which echoes out certain phase errors~\cite{supp}. Additionally, since Rb should remain in $\ket{0}$ after every pulse, we can again use flagging to increase the fidelity of the data qubits.

We benchmark our mediated gate on three-atom Cs-Rb-Cs chains, demonstrating parallel Bell state preparation of Cs mediated by Rb (Fig.~\ref{fig:cluster}b). An initial $\pi/2$-pulse puts the Cs atoms into the state $\ket{++}=(\ket{00}+\ket{01}+\ket{10}+\ket{11})/2$, and the (effective) Rb $2\pi$-pulse creates the Cs Bell state $(-\ket{00}+\ket{01}+\ket{10}+\ket{11})/2 = (\ket{1+}-\ket{0-})/\sqrt{2}$. To obtain the fidelity of this state, we perform two measurements on Cs: a population measurement in the $Z$ basis, and a coherence measurement using a global $\pi/4$-pulse to extract $2r(\vartheta)=\langle (R(\vartheta)+Z)^{\otimes 2} \rangle$. From the population and coherence measurements, we extract a Bell state fidelity of 96.7(1.7)\% after flagging and SPAM correction~\cite{supp}.

When applied to larger chains, the same pulse sequence generates a 1D cluster state on the Cs atoms. We perform this experiment on arrays of 17 Cs and 16 Rb atoms, where the Rb atoms mediate pairwise CZ gates between the data qubits (Fig.~\ref{fig:cluster}c). We can verify entanglement of this cluster state by measuring $ZXZ$ stabilizers across the array~\cite{toth2005entanglement}. To do so, we use our AOD light shifts to apply a targeted $\pi/2$-pulse prior to readout, either on all of the even-index Cs atoms or all of the odd ones; the parity of three neighboring qubits then constitutes a measurement of $ZXZ$ (or $XZX$)~\cite{supp}. Combining the even and odd datasets, we find a mean stabilizer value of 0.80(1), with every stabilizer above the threshold level of 0.5. This verifies that our cluster state is entangled with respect to any cut of the chain.

\begin{figure}
\centerline{\includegraphics[scale=1]{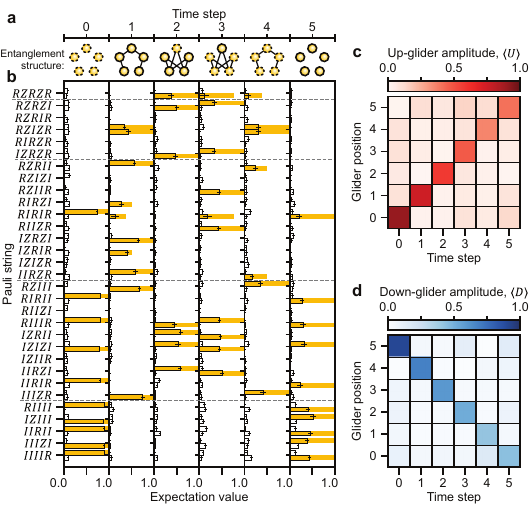}}
\caption{\textbf{Graph State QCA.} \textbf{a}, Starting with a chain of 5 Cs and 4 Rb atoms in the ground state, applying multiple Graph State unitary steps builds and unbuilds complex entanglement on Cs. These states are equivalent to graph states under single-qubit transformations of some qubits, indicated with dashed outlines. \textbf{b}, Expectation values for Pauli string operators in the Graph State QCA. For each pair of bars, the lower bar's Pauli string is indicated on the left axis, where $R$ is any unit-length combination of Pauli $X$ and $Y$~\cite{supp}; the upper bar is the same string under the replacement $R\leftrightarrow Z$. Outlined bars are experimental results, yellow bars are analytic ideal values. \textbf{c}, $U$- and \textbf{d}, $D$-operators are a subset of the aforementioned Pauli strings, corresponding to upward-moving and downward-moving gliders.}
\label{fig:graph}
\end{figure}

\section*{Graph State Automaton}

Finally, we use our novel mediated gate to implement a fundamentally quantum automaton. Our unitary step consists of a global $\pi/2$-rotation on the Cs qubits (i.e. single-qubit $\sqrt{X}$ gates), followed by parallel Rb-mediated CZ gates between each Cs pair,
\begin{equation}\label{eqn:graph_qca_step}
    U_\text{Graph} = \left(\prod_{i=1}^{N-1} \text{CZ}_{i,i+1}\right) \prod_{i=1}^N \sqrt{X}_i.
\end{equation}
We note that a single application of this unitary step is equivalent to our Bell/cluster state preparation procedure. As more unitary steps are applied, longer-range entanglement is produced, and eventually the entanglement is unwound again. The entanglement structure is equivalent to the graph states indicated in Fig.~\ref{fig:graph}a~\cite{schlingemann2001qec}. From this state generation property, we refer to this protocol as the Graph State automaton.

With the same even/odd readout technique as used for measuring the cluster state stabilizers, we obtain the expectation values of many Pauli string operators from only a few measurements (Fig.~\ref{fig:graph}b). This can be used as a proxy for the overlap with the desired state, since some operators (indicated by yellow bars) are expected to take non-zero values for a particular choice of the readout basis $R(\alpha)$~\cite{supp}. We find that the measured values of these operators are indeed noticeably larger than the rest, suggesting coherence is being preserved across multiple time steps.

A subset of the operators in Fig.~\ref{fig:graph}b act as \emph{gliders}~\cite{farrelly2020review} which remain compact while they travel across the qubit array. There is a set of upward-moving gliders, $U_i = X_{i-1} Z_i$ ($U_0 = Z_0$, $U_5 = X_5$), and a set of downward-moving gliders, $D_i = Z_i X_{i+1}$ ($D_0 = X_0$, $D_5 = Z_5$). When the unitary step is applied, the gliders are transformed as $U_\text{Graph}^\dagger U_0 U_\text{Graph} = U_1$, $U_\text{Graph}^\dagger U_1 U_\text{Graph} = U_2$, and so on, with the $D$-gliders decrementing with each step instead of incrementing. 

Experimentally, we begin evolution from the all-$\ket{0}$ state, which yields $U_0=1$ and $D_5=1$, indicating the presence of two gliders (Fig.~\ref{fig:graph}c and d). As evolution continues, we indeed observe the $U$-glider translating upward across the array, and the $D$-glider moving downward. With further evolution, we would expect the gliders to change type at the boundary (e.g. $U_\text{Graph}^\dagger U_5 U_\text{Graph} = D_5$) and continue moving in the opposite direction. Such gliders can be further understood by mapping the Graph State QCA to a free-fermion model by a Jordan-Wigner transformation~\cite{jordan1928transform, sachdev1999quantum}. In this picture, gliders emerge as the real-space image of pairs of Majorana particles which remain localized under the QCA dynamics, transporting information through the array without dispersion.

\section*{Discussion and outlook}

In this work, we have demonstrated that our dual-species platform is naturally suited for implementing quantum cellular automata, owing to the scalability of neutral-atom systems, the ease of globally-addressed interactions using laser-driven Rydberg transitions, and crucially, the ability to independently control two sets of qubits~\cite{singh2023midcircuit, anand2024dual}. With the PXP automaton, we have explored many-body dynamics of quasiparticles and generation of multi-qubit GHZ states. We further developed a mediated entangling gate, bypassing Rydberg blockade limitations to create high-fidelity Bell states and 17-qubit cluster states. This mediated gate enables access to even more QCA protocols, such as our Graph State QCA, which demonstrates complex entanglement generation and propagation of localized gliders.

While this work restricts itself to one-dimensional chains of atoms, these protocols can be straightforwardly extended to two-dimensional arrays by choosing new trapping geometries~\cite{singh2022dual} and reshaping the Rydberg lasers from focused Gaussian beams into uniform rectangular beams. This uniform intensity would also help with the coherence of our Rydberg states, as would improved intensity stability and optimal pulse control~\cite{evered2023highfidelity}. Additionally, coherence could be extended from microseconds to hundreds of milliseconds by using hyperfine clock qubits~\cite{singh2023midcircuit}; unitary steps would then be implemented as Rydberg-mediated multi-qubit gates~\cite{beterov2015rydberg, anand2024dual}.

The QCA framework presents a wide range of opportunities for exploring many-body dynamics and digital approaches to quantum simulation~\cite{cesa2026engineering}. For example, the Graph State QCA acts as a stroboscopic implementation of the kicked Ising model~\cite{prosen2007chaos, prosen2002general, kim2023evidence}; different phases of this model could be explored by tuning the ``kick" strength (given by the Cs single-qubit rotations) and the Ising interaction strength (given by the phase of our mediated controlled-phase gate) applied at each time step. Other unitary steps offer routes towards explorations of quantum chaos and quantum many-body scars~\cite{gopalakrishnan2018operator, rozon2022constructing, iadecola2020nonergodic, giudici2024unraveling}, as well as variational approaches to quantum state preparation or combinatorial  optimization \cite{wintermantel2020unitary, moll2018optimization, cerezo2021variational}. Global controls are sufficient for universal quantum computation~\cite{benjamin2001simple, levy2002universal, raussendorf2005qca, cesa2023universal}, and may enable avenues towards measurement-free error correction~\cite{fitzsimons2009fault, guedes2024qca}. Adding the mid-circuit measurement capabilities of our dual-species system~\cite{singh2023midcircuit} further unlocks the study of measurement-induced phase and entanglement transitions~\cite{iaconis2020measurement}. Thus, owing to the tunability and scalability of QCAs, the globally-controlled experimental protocols introduced here pave the way for a myriad of novel explorations in quantum computation, state preparation, and simulation.

\noindent \emph{Note added.} This manuscript is submitted simultaneously with closely related theoretical work on implementing discrete local dynamics in globally-driven dual-species arrays~\cite{cesa2026engineering}.

\section*{Acknowledgments}

We thank Kevin Singh, Bryce Gadway, and Peter Zoller for their feedback on the manuscript. We acknowledge funding from the Office of Naval Research (N00014-23-1-2540), the Air Force Office of Scientific Research (FA9550-21-1-0209), Q-NEXT, a US Department of Energy Office of Science National Quantum Information Science Research Center, the Army Research Office (W911NF2410388), the European Union’s Horizon Europe research and innovation program under Grant Agreement No. 101113690 (PASQuanS2.1), the ERC Starting Grant QARA (101041435), and the Austrian Science Fund (FWF) (DOI 10.55776/COE1). R.W. is supported by the National Science Foundation Graduate Research Fellowship under Grant No. 2140001. G.G. is supported by the European Union’s Horizon Europe program under the Marie Sk\l{}odowska Curie Action TOPORYD (101106005). T.I. is supported by NSF Grant Number DMR-2611305 and gratefully acknowledges the Kavli Institute for Theoretical Physics, which is supported by NSF grant PHY-2309135, where part of this work was performed.

\bibliography{QCA}

\end{document}



\section*{Methods}

\setcounter{figure}{0}
\renewcommand{\thefigure}{\arabic{figure}}
\setcounter{table}{0}
\renewcommand{\thetable}{\arabic{table}}
\setcounter{equation}{0}
\renewcommand{\theequation}{\arabic{equation}}

\subsection{Experimental platform}

\subsubsection{Dual-species trapping and rearrangement}

\begin{figure}[h]
\centering
\centerline{\includegraphics[scale=1]{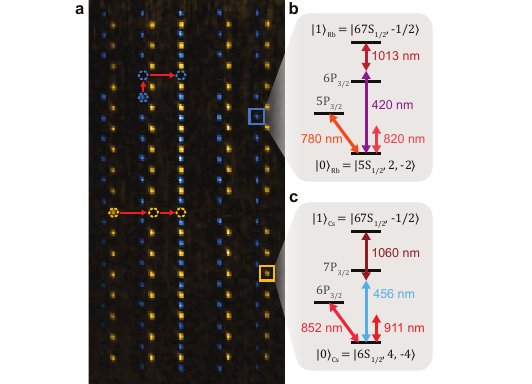}}
\caption{\textbf{a}, Background-subtracted averaged fluorescence image of our full dual-species atom array. Some sites are blacked out to show example rearrangement moves (red arrows). Middle column appears brighter because rearrangement increases the filling fraction of this column, while depleting the other columns. \textbf{b}, \textbf{c}, Relevant atomic levels and laser wavelengths for Rb and Cs, respectively.}
\label{suppfig:full_array}
\end{figure}

As in previous work~\cite{singh2023midcircuit, anand2024dual}, Rb and Cs atoms are trapped in optical tweezers generated by two SLMs, using laser light at 820.1\,nm and 911.3\,nm, respectively. Experiments are performed on a single column of atoms with Rb-Cs spacing of 5.3\,\textmu m, but to enable efficient rearrangement, four more columns are used for additional loading (Extended Data Fig.~\ref{suppfig:full_array}). The additional Cs columns are slightly horizontally offset from the Rb columns, allowing us to move atoms of either species vertically without having to route around the other species.

Rearrangement is performed using a single pair of crossed AODs at the Cs trapping wavelength (911.3\,nm), driven by the amplified outputs of an arbitrary waveform generator (AWG; Spectrum M4i6631-x8). Since this wavelength does not trap Rb as strongly as Cs, the voltage of AOD tones for Rb is increased by 20\% to compensate. Each movement during rearrangement is .5\,ms long, consisting of a .2\,ms ramp-up of the amplitude, a .1\,ms frequency chirp, and a .2\,ms ramp-down. Frequency chirps are split into two quadratic pieces, constructed such that they start at the given initial frequency, end at the target frequency, satisfy $\frac{df}{dt}=0$ at the start and end of the chirp, and maintain continuity of $\frac{df}{dt}$ between the two pieces.

After identifying atom occupation in an initial pair of fluorescence images, our rearrangement algorithm consists of three steps, performed separately on Rb and Cs arrays. First, for each empty row, one atom is taken from a neighboring row with at least two atoms. Second, the middle column is filled using atoms in the additional columns. During this step, multiple atoms are moved in parallel, such that only one move is performed per column. Finally, if smaller 1D arrays of atoms are desired, excess atoms are pulled out of the middle column, leaving at least three vacant sites between adjacent arrays. After the first round of rearrangement for both species, fluorescence images are taken again, and a second round of rearrangement attempts to correct any defects. A third set of images verifies final occupations, and subsequent experiments are performed, with data retained only if rearrangement was successful.

The average rearrangement success probability per-atom is 94\% for Rb and 92\% for Cs. On large arrays, this can still result in low data acquisition rates. For Fig.~1g, to increase our data rate, losses are determined by the final atomic image without selecting on loading. Instead, SPAM correction accounts for the finite chance of loading any particular atom.

\subsubsection{Rydberg control}

Since this work does not require ground state hyperfine qubits, negative stretched states ($m_F=-F$) were chosen over clock states ($m_F=0$). This simplifies optical pumping and Rydberg excitation, since there are fewer nearby excitation pathways to consider when applying $\sigma^-$ light to the ground state. Optical pumping and Rydberg excitation (to $\ket{67S_{1/2}, m_J=-\frac{1}{2}}$ states) are performed in a magnetic field of $B=11.8\,\text{G}$, oriented parallel to the 1D array. This field introduces Zeeman splittings, which allow us to isolate individual states. Conveniently, noting that the ground state g-factors are $g_F^\text{Rb}=\frac{1}{2}$ and $g_F^\text{Cs}=\frac{1}{4}$ and the Rydberg state g-factor is $g_J=2$, the stretched state shifts exactly match the Rydberg state shifts, making our Rydberg transitions insensitive to the magnetic field:

\begin{equation} \label{eqn:ryd_shift}
\begin{aligned}
\mu_B g_F^\text{Rb} m_F^\text{Rb} B &= \mu_B g_F^\text{Cs} m_F^\text{Cs} B = \mu_B g_J m_J B \\
&= h\times -16.5\,\text{MHz}.
\end{aligned}
\end{equation}

We access the Rydberg state via two-photon excitation, using $\sigma^-$–polarized blue light and $\sigma^+$–polarized infrared light. For Rb, 420.3\,nm light is generated by an M Squared Equinox + SolsTiS + ECD-X system, and 1013.0\,nm light is generated by a SolsTiS pumped with a Spectra-Physics Millennia eV, then amplified by a Precilasers fiber amplifier (FA-SF-1013-80-CW). For Cs, 455.7\,nm light is generated by a Vexlum VALO SHG SF system, and 1060.2\,nm light is generated by a Precilasers DBR fiber seed (YFL-SF-1059-S) with a Precilasers fiber amplifier (YFL-SF-1059-50-CW). Laser parameters and resulting Rabi frequencies are listed in Extended Data Table~\ref{tab:laser_pars}.

We have found previously~\cite{anand2024dual} that our chosen pair of Rydberg states experiences a van der Waals (vdW) interaction with coefficient $C_6=662(21)\,\text{GHz}\,\text{\textmu}\text{m}^6$. The vdW interaction is shorter-ranged than a dipole-dipole interaction, for which the next-next-nearest-neighbor interactions ($V_\text{NNNN}$) would be $1/3^3=3.7\%$ the strength of nearest-neighbor interactions ($V_\text{NN}$). By contrast, the vdW interaction has a next-nearest-neighbor ratio of $V_\text{NNN}/V_\text{NN}\approx1/2^6=1.6\%$ which allows us to better satisfy the blockade requirement $V_\text{NN}\gg\Omega\gg V_\text{NNN},V_\text{NNNN}$.

\begin{table}[!h]
    \centering
    \begin{tabular}{c | c | c | c | c}
        \textbf{Parameter} & \textbf{Rb blue} & \textbf{Rb IR} & \textbf{Cs blue} & \textbf{Cs IR} \\
        \hline
        Wavelength (nm) & 420.3 & 1013.0 & 455.7 & 1060.2 \\
        Power (W) & .018 & 8.5 & .054 & 10 \\
        Waist (\textmu m) & 60.8(1.1) & 42.7(4) & 65.9(1.8) & 43.4(1.0) \\
        \makecell{Single-photon Rabi \\ frequency (MHz)} & 109 & 120 & 124 & 209 \\
        \makecell{Intensity noise \\ (std.~dev.)} & 1.4\% & 0.3\% & 0.5\% & 0.5\% \\
        \makecell{Intermediate state \\ detuning (GHz)} & \multicolumn{2}{c|}{+2.3} & \multicolumn{2}{c}{-4.3} \\
        \makecell{Two-photon Rabi \\ frequency (MHz)} & \multicolumn{2}{c|}{2.9} & \multicolumn{2}{c}{2.9} \\
    \end{tabular}
    \caption{Typical parameters for Rydberg excitation lasers}
    \label{tab:laser_pars}
\end{table}

\subsubsection{Rydberg laser frequency noise measurements}

\begin{figure}[b]
\centering
\centerline{\includegraphics[scale=1]{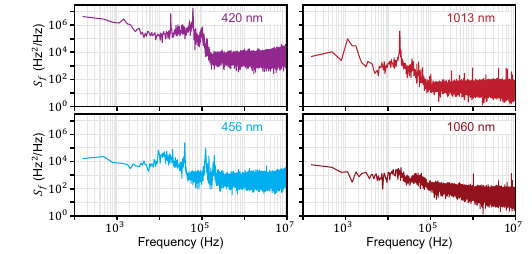}}
\caption{Frequency noise power spectral density $S_f$ for each of our four Rydberg lasers, calculated from self-heterodyne measurements on two Mach-Zehnder Interferometers with a 10\,m delay line.}
\label{suppfig:freq_noise}
\end{figure}

To better understand the noise sources in our system, two Mach-Zehnder Interferometers (MZIs) were constructed to analyze the frequency noise spectrum of our Rydberg lasers. The MZIs are entirely fiber-based, and to avoid significant wavelength-dependent losses, one MZI is configured for blue light, and the other for infrared. Each MZI consists of two fiber-coupled beamsplitters, a 10\,m delay line on one arm, a fiber-coupled AOM on the other arm, and a high-bandwidth balanced photodetector for final differential measurement (Thorlabs, PDB415A/C). Time-domain data is collected on an oscilloscope, then processed digitally (Extended Data Fig.~\ref{suppfig:freq_noise}).

For each of the four lasers, $<1$\,mW of light is fiber-coupled and sent to the respective MZI, sampled either after passing through the glass cell (for the blue lasers) or from the monitor output of the fiber amplifier (for the IR lasers). All four lasers are simultaneously locked to a single ultra-low expansion (ULE) cavity~\cite{anand2024dual}, to stabilize the laser frequency and reduce frequency noise up to the bandwidth of the lock. We find generally good performance of the Rydberg lasers, with frequency noise power spectral density $S_f$ mostly below the $10^4$\,Hz$^2$/Hz level, excluding a few resonant peaks which should not significantly couple to our Rabi driving ($\Omega\approx 3$\,MHz). The exception is our 420\,nm laser, whose increased frequency noise we attribute to declining performance of its Equinox pump laser.

\subsubsection{Automated beam alignment}
Beam pointing is stabilized via a semi-automated alignment protocol. Each Rydberg laser is positioned using a 2–axis piezo-actuated Picomotor mirror mount, and sampled beam positions are measured on a board camera (DMM 27UP031-ML). A feedback loop allows us to align each beam to a target position on the camera. At the start of each day, a 2D raster scan of the beam positions is run on all Rydberg lasers. At each beam position as aligned to on the board camera, the differential Stark shift caused by the Rydberg laser on the clock states of the atoms of that species is measured. Fitting these position-dependent Stark shifts to a Gaussian gives us the optimal beam position for the day, i.e., the position with maximal Stark shift. During the day, we periodically re-align the beams to these optimal positions. We find that without correction, the beams drift about 3\,\textmu m/day on average.

\subsubsection{AOD light shifts}

To perform a selective initialization or readout pulse, we use the rearrangement AODs to apply stationary tweezers onto atoms, shifting their Rydberg level via AC Stark shift. Since the AOD tweezers' wavelength (911.3\,nm) produces a larger shift on Cs atoms than Rb, only the Cs atoms are addressed in this manner. To minimize losses on the targeted atoms, a short ($\sim$180\,ns) pulse is used, achieved by pulsing an AOM upstream of the AOD; this AOM is otherwise used for intensity stabilization. The TTL signal for this pulse is delivered by the Rydberg AWG, ensuring synchronized timing of the AOD and Rydberg pulses.

Alignment of the AOD tweezers to the SLM-trapped Cs atoms is done by adding Zernike polynomials to the SLM hologram to maximize the AC Stark shift on the Rydberg level. After aligning the tweezers, the per-atom AC Stark shifts are measured and uniformized to within 3\% (standard deviation) by feeding back on the tone amplitudes. Finally, the laser power setpoint is swept to confirm low losses on ground state atoms. At our chosen setpoint, the AOD pulse even increases Cs atom retention, since the trapping potential catches atoms while the SLM tweezers are switched off. Ultimately, we achieve a mean shift of 35\,MHz, sufficiently large ($\gtrsim 11\,\Omega$) to suppress Rydberg excitation on targeted atoms. Fig.~1f shows a 33\% larger shift, on account of using increased laser power for uniformization purposes.

\subsection{Simulations}
\label{suppsect:simulations}

Numerical simulations were performed using the quantum trajectories method~\cite{daley2014trajectories} on systems of up to 11 atoms (6 Rb and 5 Cs). Each atom is modeled as a three-level system $\ket{0}$, $\ket{e}$, and $\ket{1}$, where $\ket{e}$ denotes the intermediate state of the two-photon Rydberg excitation $\ket{0} \leftrightarrow \ket{1}$.

The dynamics of each trajectory is computed exactly using sparse-matrix methods and includes off-resonant scattering from the intermediate state $\ket{e}$, Rydberg-state decay, laser phase noise, shot-to-shot atomic position fluctuations, and laser intensity fluctuations. Decay from $\ket{e}$ and $\ket{1}$ is assumed to occur via a single channel into ground states outside the three-level manifold. Laser phase noise is modeled as white frequency noise, with an amplitude fixed by the frequency-averaged power spectral densities shown in Extended Data Fig.~\ref{suppfig:freq_noise}. Unless otherwise specified, the simulations presented here do not include state-preparation-and-measurement (SPAM) errors. We verified separately that including such errors yields results consistent with the raw experimental data.

Since the preparation of a 17-site cluster state is beyond the reach of exact numerical simulations, this experiment was instead modeled using tensor-network methods (including three levels per atom) in the absence of noise. In this case, deviations from an ideal cluster state arise from the finite Rydberg blockade strength, next-nearest-neighbor interactions, and the finite detuning of the intermediate state.

\subsection{PXP QCA, quasiparticles}

\begin{figure}
\centering
\centerline{\includegraphics[scale=1]{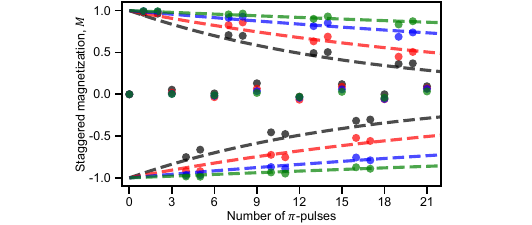}}
\caption{Comparison of the decay in the staggered magnetization $M$ for experimental data (black), a noiseless simulation (green), a noisy simulation (blue), and a noisy simulation with a small detuning error, $\Delta\approx\Omega/10$ (red). Data are fit to the classical result with an exponential damping term, and dashed lines are the exponential part of this fit.}
\label{suppfig:magnetization}
\end{figure}

The unitary step for the PXP automaton is given in Eq.~1 of the main text, under the assumption of infinite nearest-neighbor interactions, and vanishing interactions beyond the nearest-neighbor. A single Floquet step is defined by two pulses, $U_F=U_\text{odd} U_\text{even}$, and this automaton features a system-size-independent cycle that repeats every 3 Floquet steps: 

\begin{equation} \label{eqn:pxp_vac}
\begin{aligned}
\ket{\text{vacuum}} \equiv \ket{0000...} &\xrightarrow[]{U_F} \\
\ket{\mathbb{Z}_2^\text{(even)}} \equiv \ket{1010...} &\xrightarrow[]{U_F} \\
\ket{\mathbb{Z}_2^\text{(odd)}} \equiv \ket{0101...} &\xrightarrow[]{U_F} \ket{\text{vacuum}}
\end{aligned}
\end{equation}

The van der Waals interaction perturbs the PXP unitary step, causing leakage out of the states in this period–3 cycle~\cite{iadecola2020nonergodic}. To estimate how much of our population damping (e.g. in Fig.~1g and Fig.~2a) comes from these interactions versus noise, we perform simulations of the vacuum evolution on an 11–atom chain, and compare against experimental data (Extended Data Fig.~\ref{suppfig:magnetization}). We track the staggered magnetization at each time step, $M = \langle n \rangle_\text{Rb} - \langle n \rangle_\text{Cs}$, where the subscripts indicate that the average is taken over only the atoms of one particular species. We see that with just the van der Waals interaction, the magnetization decays on a timescale of 141(10) $\pi$–pulses, although at longer timescales than presented here one would expect coherent revivals rather than exponential decay~\cite{iadecola2020nonergodic}. Adding our noise model to the simulation reduces the decay timescale to 68(5) pulses, and experimentally we observe 17(1) pulses. To better understand our experimental results, we run a final simulation with the same noise model, but adding a constant detuning error equal to the van der Waals next-nearest-neighbor interaction strength, $\sim$0.2\,MHz for Cs and $\sim$0.3\,MHz for Rb; the decay timescale in this case is 31(2) pulses. Although our laser calibrations account for the van der Waals detuning, it is plausible that slow drifts of the AC stark shift (e.g. arising from slight laser pointing errors) could introduce a similar error after enough time.

 Considering the vacuum orbit states of Eq.~\ref{eqn:pxp_vac}, when different sub-arrays follow different vacuum configurations, a domain wall quasiparticle exists at their interface. For example, the state $\ket{10100001010}$ has a $\mathbb{Z}_2^\text{(even)}$ configuration on the left qubits, a vacuum configuration in the middle, and a $\mathbb{Z}_2^\text{(odd)}$ configuration on the right, making two quasiparticles. Following~\cite{wilkinson2020exact, iadecola2020nonergodic}, quasiparticles can be conveniently identified in each shot of the experiment by looking for one of three bitstrings across four adjacent qubits: $0001$, $1000$, or $1001$. These bitstrings are mutually exclusive, so only one quasiparticle can be observed at a given location at a time. At the edges of the chain, there is only a single quasiparticle type, with a three-qubit bitstring identifier: $001$ for the left end, or $100$ for the right end. Writing $P_j=\ket{0}\bra{0}_j$ and $n_j = \ket{1}\bra{1}_j$, for a system with $L$ qubits we can define $L-1$ quasiparticle density operators $q_j$,
\begin{equation} \label{eqn:pxp_qj}
\begin{gathered}
q_1 = P_0 P_1 n_2, \\
q_j = P_{j-2} P_{j-1} P_{j} n_{j+1} + n_{j-2} P_{j-1} P_{j} P_{j+1} \\ + n_{j-2} P_{j-1} P_{j} n_{j+1}, \\
q_{L-1} = n_{L-3} P_{L-2} P_{L-1},
\end{gathered}
\end{equation}
from which we calculate the total quasiparticle number,
\begin{equation} \label{eqn:pxp_Q}
Q = \sum_{j=1}^{L-1} q_j.
\end{equation}

In Fig.~2e, we omit multiple-of-three time steps because these are more sensitive to errors in the quasiparticle number. In the case where the rotation angle is exactly $\pi$, the state of the system at these times should approximately be the 0–quasiparticle state $\ket{\text{vacuum}}$. However, if a single atom is accidentally lost, we will instead measure $\ket{0...010...0}$, a two-particle state. As a result, we observe higher quasiparticle numbers at these times, and we omit them for visual clarity.

\subsection{GHZ states}

Greenberger-Horne-Zeilinger (GHZ) states are generated naturally by the PXP automaton upon initialization of a single atom in the $\ket{+}$ state. Generation of the GHZ state can be understood in terms of quasiparticles, as described in the main text. Alternatively, one can simply follow the evolution pulse-by-pulse. For example, showing the first few $\pi$-pulses, with the entangled atoms specified by color:
\begin{equation} \label{eqn:ghz_evo}
\begin{aligned}
\ket{00000\textcolor{cs}{0}00000} &+ \ket{00000\textcolor{cs}{1}00000} \xrightarrow[]{\pi_\text{Rb}} \\
\ket{1010\textcolor{rb}{1}\textcolor{cs}{0}\textcolor{rb}{1}0101} &+ \ket{1010\textcolor{rb}{0}\textcolor{cs}{1}\textcolor{rb}{0}0101} \xrightarrow[]{\pi_\text{Cs}} \\
\ket{1010\textcolor{rb}{1}0\textcolor{rb}{1}0101} &+ \ket{1010\textcolor{rb}{0}0\textcolor{rb}{0}0101} \xrightarrow[]{\pi_\text{Rb}} \\
\ket{0000\textcolor{rb}{0}0\textcolor{rb}{0}0000} &+ \ket{0000\textcolor{rb}{1}0\textcolor{rb}{1}0000} \xrightarrow[]{\pi_\text{Cs}} \\
\ket{010\textcolor{cs}{1}\textcolor{rb}{0}\textcolor{cs}{1}\textcolor{rb}{0}\textcolor{cs}{1}010} &+ \ket{010\textcolor{cs}{0}\textcolor{rb}{1}\textcolor{cs}{0}\textcolor{rb}{1}\textcolor{cs}{0}010} \xrightarrow[]{\text{...}} \\
\end{aligned}
\end{equation}

Fidelity measurements are performed on the entangled atoms as indicated above, after postselecting on the auxiliary atoms being in the ground state (for single-species time steps only). To account for the possibility of a global phase shift on the qubits, we define the following class of $N$–atom GHZ-like states,
\begin{equation}
    \ket{\text{GHZ}(\phi)} = \frac{\ket{0}^{\otimes N} + e^{i\phi}\ket{1}^{\otimes N}}{\sqrt{2}},
\end{equation}
which are equivalent to the standard GHZ state up to uniform local phase gates, $\ket{\text{GHZ}(\phi)} = \left(e^{-i\phi Z/2N}\right)^{\otimes N}\ket{\text{GHZ}(0)}$. Given the experimentally prepared state $\rho$, we are interested in the generalized GHZ fidelity
\begin{equation}
    F_\text{GHZ} = \max_\phi\langle \text{GHZ}(\phi) | \rho | \text{GHZ}(\phi) \rangle.
\end{equation}
We can extract this fidelity via global measurements by noting that it can be bounded as
\begin{equation}
    F_\text{GHZ} \geq Q-\frac{1}{2} +\frac{1}{2}\max_\phi B(\phi),
\end{equation}
where we defined $Q=\langle 0^{\otimes N} | \rho | 0^{\otimes N}\rangle + \langle 1^{\otimes N} | \rho | 1^{\otimes N}\rangle$ and $B(\phi) = \langle R^{\otimes N}(\phi)\rangle$, with $R(\alpha) = \cos(\alpha)X+\sin(\alpha)Y$ a generic operator in the equatorial plane of the Bloch sphere. We can measure $Q$ by simply reading out the atomic populations in the $Z$ basis, and $B(\phi)$ by first applying a rotation of an angle $\pi/2$ with phase $\phi-\pi/2$ before the readout. Finally, the maximum of $B(\phi)$ is estimated by fitting the data to a cosine with period $2\pi/N$, and the generalized GHZ fidelity is calculated (Fig.~3c).

\begin{figure*}[!t]
\centering
\centerline{\includegraphics[scale=1]{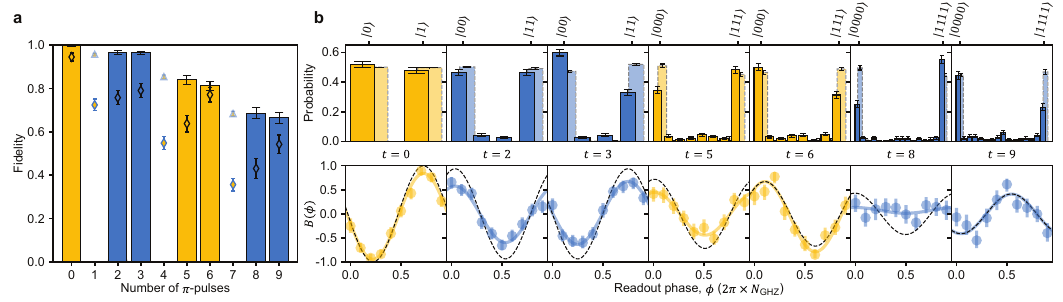}}
\caption{\textbf{a}, Comparison between numerically simulated and experimentally measured GHZ state fidelities in the PXP automaton. The vertical bars show the simulated fidelities including noise and intermediate state scattering. The colored diamonds are the experimentally determined fidelities as shown in Fig.~3 of the main text. For the two-species GHZ states, we additionally show the simulated state fidelity as obtained directly from the wavefunction overlap (light-colored triangles). The error bars on the noisy simulations denote the $1\sigma$ standard deviation across 100 trajectories. \textbf{b}, Detailed population (top row) and parity (bottom row) results for the time steps shown on the left. For the population plots, dark bars show experimental measurements, and the light bars with dashed outline are simulation results. For the parities, colored data points are experimental measurements, the colored solid lines are sinusoidal fits to the data, and the black dashed lines are simulation results. Note that the phase offset of the simulated parity sweep has been calibrated to the measured offset for visual clarity.}
\label{suppfig:GHZ_simulated}
\end{figure*}

For the time steps at which both species are entangled (i.e. time steps 1, 4, and 7; striped bars in Fig.~3c), the GHZ state is of the form $\ket{010}+\ket{101}$ rather than $\ket{000}+\ket{111}$. Extracting $Q$ is still possible by measuring the population of the corresponding states, but $B(\phi)$ cannot be measured because of the Rb-Cs blockade on the pre-readout pulse. To obtain an estimate of the fidelity, we use the best parity of any subsequent measurement, $B_{t_0} = \underset{t>t_0}{\max} (B_t )$, under the assumption that additional pulses do not increase the coherence of the state. While this does not rigorously calculate the GHZ state fidelity, comparison with numerically simulated fidelities indicates that this strategy consistently provides a conservative lower bound (Extended Data Fig.~\ref{suppfig:GHZ_simulated}a, light-colored triangles)

\subsubsection{Comparison with numerical simulations}

To obtain further insight on the relative influence of certain error sources in our experiment on the state fidelity, we compare our experimental results with numerical simulations. As described in Section~\ref{suppsect:simulations}, we simulate our PXP QCA protocol using a quantum trajectories algorithm on 11 atoms (6 Rb and 5 Cs). Extended Data Fig.~\ref{suppfig:GHZ_simulated} shows the obtained state fidelities as well as raw populations and parity sweeps for each time step, evaluated exactly as for the experimental data. In comparison to the measured preparation fidelities, we find that the simulation predicts a decay in fidelity over a similar timescale as observed in the experiment. However, the simulation predicts consistently higher fidelities, which hints to additional noise sources that are not captured by the simulation, or the presence of small detuning errors (cf. Extended Data Fig.~\ref{suppfig:magnetization}) which we found to have strong influence on the fidelity.

\subsection{Mediated gate}

Here, we present the details of our mediated gate (MG). As introduced in the main text, we can implement MGs between two data atoms (initially in any state) when an auxiliary qubit is placed between them, and initialized in $\ket{0}$. Then, the MG proceeds by applying a pulse sequence that, by driving both data and auxiliary, applies an entangling gate on the data, while the auxiliary returns deterministically to $\ket{0}$.

\subsubsection{Mediated gate protocol}
\label{suppsect:mediated_gate_protocol}

To design our MG, we consider a three atom system, composed of two data atoms and one auxiliary atom between them. The distances are tuned such that the auxiliary atom interacts strongly with both data atoms, while the latter are far outside each other's blockade radii. We are then interested in engineering the gate
\begin{equation}
    \text{CZ} = \ket{11}\bra{11} +  \ket{01}\bra{01} +  \ket{10}\bra{10} -  \ket{00}\bra{00}
\end{equation}
on the data, using the auxiliary as a mediator of the interaction.

In the ideal blockade regime, a protocol to achieve this is the following: one can simply drive a $2\pi$ pulse on the auxiliary species, which, in the absence of interactions, would induce a closed trajectory in the Bloch sphere, resulting in a $-1$ global geometrical phase. Due to the data-auxiliary blockade, this only occurs if the data atoms are initially in $\ket{0}$, making the phase conditional and resulting in a CZ gate on the data qubits (Fig.~4a).

However, for high fidelity gates the protocol above is limited by finite interspecies interactions and nonzero intraspecies interactions, as well as systemic phase noise. Indeed, while we can achieve high-quality inter-species blockade, we find non-negligible interactions between the two data atoms. We thus design a more sophisticated protocol, incorporating a built-in echo technique to cancel such effects, which can be optimized directly on the device. The pulse sequence proceeds in three steps:

(i) We first apply a pulse to Rb (which is initialized in $\ket{0}$) with detuning $\Delta$ and duration $2\pi/\sqrt{\Omega^2 + \Delta^2}$, such that it traverses a closed trajectory in the Bloch sphere, mapping it back to $\ket{0}$. Due to the blockade interaction with Cs, this effectively results in a diagonal gate on the data atoms:

\begin{equation}
    U_\Delta = \Big[e^{i(\phi-\Phi)Z/4}\otimes e^{i(\phi-\Phi)Z/4} \Big] e^{i(\phi + \Phi)ZZ/4};
\end{equation}

\noindent here, $\Phi = \pi(1 - \Delta/\sqrt{\Omega^2+\Delta^2})$ is uniquely determined by our choice of detuning, while $\phi$ is an \emph{unknown} phase that encompasses the long-range interaction of the two Cs atoms and any additional global phase error.

(ii) For the second step, we simply apply a resonant $\pi$-pulse to Cs. Since in the previous step Rb was mapped back to $\ket{0}$, there is no blockade constraint in action; the resulting operation simply flips the Cs states, thereby implementing an $X\otimes X$ gate on the data atoms.

(iii) Finally, we repeat the very same driving as in step (i), with the same choice of $\Delta$. Thus, we apply $U_\Delta$ a second time. The resulting operation therefore reads:
\begin{equation}\label{eqn:V_delta}
    V_\Delta = U_\Delta (X\otimes X) U_\Delta = (X\otimes X)\Big[e^{i\alpha Z / 2}\otimes e^{i\alpha Z / 2} \Big] \text{CZ}(2\alpha),
\end{equation}
where we defined the generalized two-qubit phase gate $\text{CZ}(\xi) = I - (1-e^{i\xi}) \ket{00}\bra{00} $. We stress again that the angle $\alpha = \phi+\Phi$ is unknown. However, we can tune our control knob $\Delta$, such that $\alpha = \pi/2$, resulting in a CZ gate (up to local corrections, which can be handled in the compilation stage). After applying the gate, the auxiliary atom should remain in $\ket{0}$, and can be traced out or measured to remove some sources of error. To find the optimal $\Delta$, we proceed as in the next section.

\subsubsection{Gate optimization and Bell state preparation}

Our goal is to tune $\Delta$ such that $V_\Delta$ is a maximally entangling gate. 
To do so, we use the following procedure: (i) Initialize the data atoms in $\ket{++}$ with a global Cs $\pi/2$-pulse; (ii) Apply $V_\Delta$; (iii) Apply a second $\pi/2$-pulse, about the original axis; (iv) Choose the value of $\Delta$ that minimizes $|\langle Z_1 \rangle \langle Z_2 \rangle|$. Since the ideal value ($|\langle Z_1 \rangle \langle Z_2 \rangle|=0$) is reached when the unknown phase $\phi$ is exactly compensated by $\Phi(\Delta)$ to provide a maximally entangled gate, this procedure allows us to identify the optimal detuning $\Delta$. Experimentally, we obtain the value $\Delta\approx1.25\,\Omega$.

Given the experimentally optimized detuning $\Delta$, we then use our protocol to prepare a Bell state as described in the main text. To benchmark our state preparation, we define states of the generic form
\begin{equation}
    \ket{C(\varphi)} = \frac{e^{-2i\varphi}\ket{00} + \ket{01}+\ket{10} - e^{2i\varphi}\ket{11}}{2}. 
\end{equation}
These are locally equivalent to the two-qubit cluster state, $\ket{C(\varphi)} = e^{-i\varphi Z}\otimes e^{-i \varphi Z} (\ket{0+}+ \ket{1-})/\sqrt{2}$, which is related to a standard two-qubit Bell pair by a local Hadamard gate (on any of the two qubits), $\ket{C(0)} = H\otimes I(\ket{00}+\ket{11})/\sqrt{2}$. Given an experimentally generated state $\rho$, we can define the fidelity 
\begin{equation}
    F_C = \max_\varphi \langle C(\varphi) | \rho | C(\varphi) \rangle. 
\end{equation}
To extract this fidelity through uniform operations only, we find it convenient to rewrite this equation as
\begin{equation}
    F_C = \frac{1-P}{4} + \max_\vartheta W(\vartheta),
\end{equation}
where we have $4W(\vartheta) = r(\vartheta) + r(\vartheta + \pi/2) + r(\vartheta - \pi/2) - r(\vartheta+ \pi)$. Herein, we defined the uniform observables $P = \langle Z^{\otimes 2}\rangle$ and $2r(\vartheta) = \langle (R(\vartheta)+Z)^{\otimes 2} \rangle$, with $R(\vartheta) = \cos(\vartheta)X+\sin(\vartheta)Y$. We can measure $r(\vartheta)$ by simply applying a $\pi/4$ pulse with phase $\vartheta-\pi/2$ before the readout.

To estimate $\max_\vartheta W(\vartheta)$, we fit our measured $r(\vartheta)$ to the function
\begin{equation}
    r_\text{fit}(\vartheta;A,B,\vartheta^*) = 2AB (B\sin^2(\vartheta-\vartheta^*) + 2\cos(\vartheta-\vartheta^*)),
\end{equation}
where the parameters $A$ and $B$ account for depolarization and correlated phase errors. For the maximally entangling gate, corresponding to $\alpha = \pi/2$ in Methods Eq.~\ref{eqn:V_delta}, the angle maximizing $W(\vartheta)$ is expected to be fixed to $\vartheta^* = 3\pi/4$ \emph{independently} of the unknown phase $\phi$. The fit to our experimental data (Fig.~4b) yields a value of $\vartheta^*=0.718(4)\pi$.

\subsubsection{Mediated gates in a many-body atom array}

The mediated-gate protocol extends directly to our larger dual-species 1D arrays. Specifically, as introduced in the main text, if we apply the sequence above when the auxiliary atoms are initialized in $\ket{0}$, we obtain CZ-like gates between each pair of nearest-neighbors of the data species. However, to understand the discussion below, it is useful to highlight here a key boundary effect. By implementing the sequence of Section~\ref{suppsect:mediated_gate_protocol} with the optimal detuning (i.e. $\alpha=\pi/2$) on an array featuring $N$ data atoms, the resulting unitary reads  \begin{equation}\label{eqn:many_body_CZ_layer}
    V_\text{array} = \left[S_1S_N\prod_{i=1}^N Z_i\right] \prod_{i=1}^{N-1} \text{CZ}_{i,i+1},
\end{equation}
where $S=e^{-i\pi Z/4}$. That is, we obtain a layer of nearest-neighbor CZ gates, up to local corrections; but crucially, this correction is different for the boundary atoms, with an additional $\pi/2$ rotation around the $\hat{z}$ axis. 

\begin{figure}[h]
\centering
\centerline{\includegraphics[scale=1]{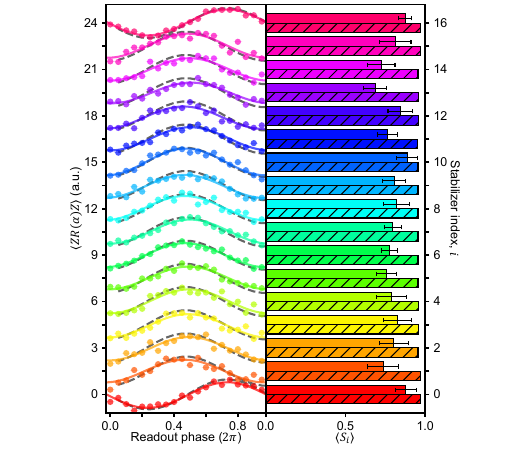}}
\caption{Complete stabilizer measurements of the 17-atom cluster state, colored by the stabilizer index $i$. Solid lines are fits to cosines, and the maximum amplitude of this cosine gives the stabilizer value $\langle S_i \rangle$ (right, solid bars; also Fig.~4c). Noiseless simulation data is shown for comparison (left, dashed curves; right, hatched bars).}
\label{suppfig:cluster_full}
\end{figure}

\subsubsection{Cluster state}

Having established our mediated CZ gate, we now describe the creation and measurement of a one-dimensional cluster state. Preparing a cluster state is straightforward~\cite{raussendorf2003mbqc}: with all data qubits initialized in $\ket{+}$ (e.g. by a global $\pi/2$-pulse), a CZ gate is applied between every adjacent pair of qubits. We perform all of these CZ gates in parallel, by placing an auxiliary (Rb) qubit between each pair of data (Cs) qubits. This same approach could be used on a two-dimensional atom array to generate a two-dimensional cluster state, which is known to be a universal resource for measurement-based quantum computation~\cite{raussendorf2003mbqc}.

The cluster state is stabilized by the operators $S_i = Z_{i-1} X_i Z_{i+1}$ ($S_0 = X_0 Z_1$, $S_{L-1} = Z_{L-2} X_{L-1}$), and measuring the expectation value of these operators is sufficient to verify entanglement across each pair of atoms in the cluster state~\cite{toth2005entanglement}. Conveniently, all $L$ stabilizers can be extracted in two sets of measurements, by measuring the even qubits in the $Z$ basis and the odds in the $X$ basis, then vice versa.

Experimentally, stabilizer data is collected by: (i) generating a cluster state on 17 Cs atoms; (ii) using our light shift beams to shift the even or odd atoms out of resonance; and (iii) applying a $\pi/2$-pulse about the $\alpha-\pi/2$ axis prior to readout. To make our data more robust against laser drifts and to capture the phase shift of the boundary stabilizers, we perform measurements in the entire $XY$ plane by measuring $R(\alpha) = \cos{(\alpha)} X + \sin{(\alpha)} Y$; we then fit the data to the theoretically predicted sinusoidal profile as a function of $\alpha$, and report the peak amplitude. Site-by-site results are shown in Extended Data Fig.~\ref{suppfig:cluster_full}, wherein we clearly see that the fits for the two edge operators are shifted by $\pi/2$ relative to the bulk, as expected from the boundary correction in Methods Eq.~\ref{eqn:many_body_CZ_layer}.

\subsection{Graph State Automaton}

The unitary step for the Graph State Automaton nominally consists of applying a global Hadamard gate followed by a global CZ between each neighboring qubit pair. This QCA is related to the well-known kicked-Ising model~\cite{prosen2007chaos,prosen2002general}, which is a minimal, yet non-integrable setting featuring paradigmatic non-ergodic properties such as quantum chaos. To benchmark our capability of implementing such a QCA, we tune it to an integrable Clifford regime, which allows efficient probing of the dynamics via simple products of Pauli measurements. At this integrable point, the model can be mapped to free fermions, and displays gliders corresponding to pairs of Majorana particles that jointly propagate through the array. As described in the main text, in real space this corresponds to the propagating Pauli operators shown in Figs.~5C and D.     

Experimentally, our unitary step consists of a global $\pi/2$ rotation followed by a mediated entangling gate. As specified by Methods Eq.~\ref{eqn:many_body_CZ_layer}, the edge atoms experience an additional phase shift of $\pi/2$ relative to the bulk. To account for this asymmetry and to observe the glider propagation, we therefore shift the laser phase between the first and the second $\pi/2$-pulses by $\pi/2$ . Up to 5 unitary steps are applied to a 5-Cs (and 4-Rb) array, constituting a full cycle of the graph-like states depicted in Fig.~5a. We note that the final time step ($t=5$) was measured on a separate day with a reduced number of $\alpha$ samples compared to the earlier steps.

Measurement of these states uses the same procedure as the cluster state stabilizer measurements: every other Cs atom is shifted out of resonance, and a $\pi/2$-pulse about the $(\alpha-\pi/2)$–axis is applied to the non-shifted atoms. As we sweep $\alpha$, we analytically predict a certain functional dependence of the readout, and in some cases this function reaches a value of 1, meaning this Pauli string is a stabilizer of the state for this choice of $R(\alpha)$. The expected non-zero operators at each time step are listed in Extended Data Table~\ref{tab:graph_ops}, along with their fitting functions. For unlisted operators, and those whose fit function is listed as ``N/A", expectation values are not predicted to vary with $\alpha$.

\begin{figure}[h]
\centering
\centerline{\includegraphics[scale=1]{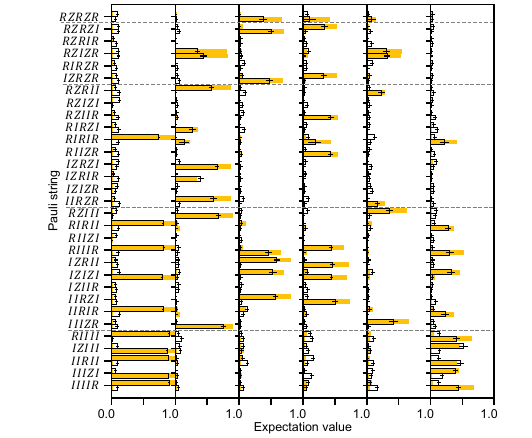}}
\caption{Comparison of experimental Graph State QCA operator values (outlined bars; from Fig.~5b) and values extracted from our noisy simulation (yellow bars).}
\label{suppfig:graph_sim}
\end{figure}

Fig.~5 reports the maximum absolute value of each fit, $|A|+|C|$, or if there is variation with $\alpha$, the absolute value of the mean of the data. Extended Data Fig.~\ref{suppfig:graph_sim} replicates the experimental data from Fig.~5, but compares against values extracted from our noisy simulation, rather than the ideal analytic values. We see significantly better agreement, though we note there is still some unaccounted loss in contrast.

\begin{table*}
    \centering
    \begin{tabular}{c | c | c | c }
        \textbf{Time step} & \textbf{Pauli string operators} & \textbf{Fit function} & \textbf{Ideal maximum} \\
        \hline
        $t=0$ & \makecell{$IIIIZ$, $IIIZI$, $IIZII$, $IZIII$, $ZIIII$, \\ $IIZIZ$, $IZIZI$, $ZIIIZ$, $ZIZII$, $ZIZIZ$} & N/A & $1$ \\
        \hline
        \multirow{4}{*}{$t=1$} &  $IIIZR$, $RZIII$, $IIZRZ$, $IZRZI$, $ZRZII$ & $A\cos{(\phi-\phi_0)}+C$ & $1$ \\
        & $RZIZR$ & $A\cos{(2(\phi-\phi_0))}+C$ & $1$ \\
        & $IZRIR$, $RIRZI$ & $A\cos{(2(\phi-\phi_0))}+C$ & $1/2$ \\
        & $RIRIR$ & $\frac{A3\sqrt{3}}{2}\cos^2{(\phi-\phi_0)}\sin{(\phi-\phi_0)}+C$ & $2/(3\sqrt{3})$ \\
        \hline
        \multirow{3}{*}{$t=2$} & $IIZRI$, $IRZII$ & $A\cos{(\phi-\phi_0)}+C$ & $1$ \\
        & $IRIRI$, $RIIIR$, $IZRZR$, $RZRZI$ & $A\cos{(2(\phi-\phi_0))}+C$ & $1$ \\
        & $RZRZR$ & $A\cos^3{(\phi-\phi_0)}+C$ & $1$ \\
        \hline
        \multirow{4}{*}{$t=3$} & $IZIZI$, $ZIIIZ$ & N/A & $1$ \\
        & $IIRZI$, $IZRII$ & $A\cos{(\phi-\phi_0)}+C$ & $1$ \\
        & $RIIZR$, $RZIIR$, $IRZRZ$, $ZRZRI$ & $A\cos{(2(\phi-\phi_0))}+C$ & $1$ \\
        & $RIRIR$, $RZRZR$ & $\frac{A3\sqrt{3}}{2}\cos^2{(\phi-\phi_0)}\sin{(\phi-\phi_0)}+C$ & $4/(3\sqrt{3})$ \\
        \hline
        \multirow{4}{*}{$t=4$} & $IIIRZ$, $ZRIII$ & $A\cos{(\phi-\phi_0)}+C$ & $1$ \\
        & $ZRIRZ$, $RZIZR$ & $A\cos{(2(\phi-\phi_0))}+C$ & $1$ \\
        & $IIRZR$, $RZRII$ & $A\cos{(2(\phi-\phi_0))}+C$ & $1/2$ \\
        & $RZRZR$ & $\frac{A3\sqrt{3}}{2}\cos^2{(\phi-\phi_0)}\sin{(\phi-\phi_0)}+C$ & $2/(3\sqrt{3})$ \\
        \hline
        \multirow{3}{*}{$t=5$} & $IIIIR$, $IIIRI$, $IIRII$, $IRIII$, $RIIII$ & $A\cos{(\phi-\phi_0)}+C$ & $1$ \\
        & $IIRIR$, $IRIRI$, $RIIIR$, $RIRII$ & $A\cos{(2(\phi-\phi_0))}+C$ & $1$ \\
        & $RIRIR$ & $A\cos^3{(\phi-\phi_0)}+C$ & $1$ \\
    \end{tabular}
    \caption{Measured operators with non-zero expectation value, and their corresponding fit functions, for the Graph State Automaton}
    \label{tab:graph_ops}
\end{table*}

\subsection{Correction of state preparation and measurement (SPAM) errors}

In this section, we discuss the different error sources due to state preparation and readout that are present in our experiment, and how we correct for them. Each parameter is calibrated experimentally using the specific protocol indicated below. Extended Data Table~\ref{tab:spam_parameters} presents an overview of the resulting parameters that are used for SPAM correction throughout the manuscript. Unless otherwise stated, all data presented in the main text is corrected for SPAM errors to faithfully assess the performance of the implemented operations.

\subsubsection{Error sources and calibration methods}

\noindent\textbf{Initial state preparation via optical pumping}

After loading the tweezers and rearrangement, the atoms are in a statistical mixture of hyperfine states. To correctly initialize both species in the respective ground states $\ket{0}_\mathrm{Cs} = \ket{F=4, m_F=-4}$ and $\ket{0}_\mathrm{Rb} = \ket{F=2, m_F=-2}$, we perform optical pumping using a circularly ($\sigma^-$) polarized laser beam resonant with the $\ket{F=2}\to\ket{F'=2}$ (Rb), and $\ket{F=4}\to\ket{F'=4}$ (Cs) transitions. Additionally, a repumper is present to remove population from the $\ket{F=1}_\mathrm{Rb}$ and $\ket{F=3}_\mathrm{Cs}$ states. The fidelity of the optical pumping step can be estimated by measuring the rate of population transfer into (\textit{pumping rate}) and out of (\textit{depumping rate}) the target state~\cite{walker2012entanglement}. These rates translate into the pumping fidelity via $\eta = 1-\tau_\mathrm{pump}/\tau_\mathrm{depump}$.

For Rb, we obtain a pumping time constant $\tau_\mathrm{pump}^\mathrm{Rb} = \SI{0.180(14)}{ms}$, and a depumping time constant $\tau_\mathrm{depump}^\mathrm{Rb} = 31.7(1.6)\,\mathrm{ms}$, corresponding to a pumping fidelity of $\eta^\mathrm{Rb}=99.43(5)\,\%$. For Cs, we obtain a pumping time constant $\tau_\mathrm{pump}^\mathrm{Cs} = \SI{0.079(2)}{ms}$, and a depumping time constant $\tau_\mathrm{depump}^\mathrm{Cs} = \SI{8.1(3)}{ms}$, corresponding to a pumping fidelity of $\eta^\mathrm{Cs}=99.03(5)\,\%$. As the pumping fidelities are directly computed from time constants, they do not require correction for other SPAM errors.

\noindent\textbf{Imaging fidelity -- Discrimination}

Due to a finite signal-to-noise ratio in the fluorescence images, the signal sum distributions of occupied and empty tweezers, according to which the presence of an atom is detected, exhibit a non-vanishing overlap. The overlap gives rise to a small probability of false positives and false negatives, which we can directly estimate from the ratio between the overlap area to the total area under the distribution. According to this, we extract a false positive rate and a false negative rate of $F_\mathrm{p}^\mathrm{Cs} = 0.63(15)\,\%$, $F_\mathrm{n}^\mathrm{Cs} = 0.76(18)\,\%$ and $F_\mathrm{p}^\mathrm{Rb} = 0.47(31)\,\%$, $F_\mathrm{n}^\mathrm{Rb} = 0.58(36)\,\%$, respectively. These rates do not need to be corrected for other error sources as they are directly obtained from averaged exposures.

\noindent\textbf{Imaging fidelity -- Survival probability}

During the imaging, the atoms scatter photons from the near-resonant optical molasses, which creates a finite loss probability as a result of the imaging process. This is particularly relevant for the image that is taken right before the experimental sequence, according to which we determine the initial occupation. To quantify the probability that an atom survives the imaging, we record two subsequent images without any further operations in between, and determine the fraction of atoms lost between the images. From this, we obtain survival probabilities of $F_\mathrm{Rb} = 98.7(6)\,\%$ and $F_\mathrm{Cs} = 97.2(2)\,\%$, respectively. The survival probabilities are corrected for image discrimination errors as quantified by the previously measured false positive and negative rates.

\noindent\textbf{Ground state detection fidelity}

As the traps are switched off during the Rydberg sequence, the atoms undergo a thermal random walk that causes ground state atoms to be recaptured with a less-than-unity probability that depends on the trap-off duration (drop time $t_\mathrm{drop}$). This is quantified using a drop-and-recapture experiment, where we switch off the traps for a varying duration before recapturing without performing any Rydberg operations in between. Such an experiment is characterized by two features: Firstly, for zero drop time ($t_\mathrm{drop} = 0$), we find a non-vanishing baseline loss that is given by the finite vacuum lifetime and, without corrections, by other static SPAM errors. Secondly, as a function of the drop time there is a monotonically increasing excess loss due to the thermal atomic motion in free flight. For a specific drop time, we identify the ground state detection infidelity with the total observed loss in this experiment. For instance, for the two-atom Bell state measurement in Fig.~4a of the main text, a drop time $t_\mathrm{drop} = \SI{1.5}{\micro s}$ was chosen. This limits the fidelity of detecting ground state atoms to $D_\mathrm{g}^\mathrm{Rb} = 96.6(5)\,\%$ for Rb, and $D_\mathrm{g}^\mathrm{Cs} = 95.9(3)\,\%$ for Cs, respectively. These detection fidelities have to be corrected for pumping infidelity, discrimination errors as well as imaging survival.

\noindent\textbf{Rydberg state detection fidelity}

Due to the finite lifetime of a Rydberg state [$\sim\mathcal{O}(\SI{100}{\micro\second})$], there is a small probability that a Rydberg state decays before leaving the recapture area, which causes an improper detection as a ground state atom. To estimate this probability, we perform single-atom Rabi oscillations between ground and Rydberg states. The sinusoidal oscillations exhibit a damping that is well-described by a single exponential envelope, with a maximum amplitude that is constrained by the Rydberg detection fidelity~\cite{deleseleuc2018analysis}. We can hence use the value of the envelope function extrapolated to $t_\mathrm{osc} = 0$ as an estimate of the Rydberg detection fidelity. According to this, we obtain uncorrected Rydberg detection fidelities of $D_\mathrm{r}^\mathrm{Cs} = 99(1)\,\%$ and $D_\mathrm{r}^\mathrm{Rb} = 94(1)\,\%$. The Rydberg detection fidelities are corrected for all previously calibrated error sources.

\begin{table}[t]
    \centering
    \begin{tabular}{c|c|c|c}
        \textbf{Parameter} & \textbf{Species} & \textbf{Raw} & \textbf{Corrected}\\
        \hline
        \multirowcell{2}{Pumping fidelity $\eta$}  & Cs & 0.9903(5) & -\\
        & Rb & 0.9943(5) & -\\
        \multirowcell{2}{Imaging false positive $F_\mathrm{p}$} & Cs & 0.0063(15) & -\\
        & Rb & 0.0047(31) & -\\
        \multirowcell{2}{Imaging false negative $F_\mathrm{n}$} & Cs & 0.0076(18) & -\\
        & Rb & 0.0058(36) & -\\
        \multirowcell{2}{Survival probability $S$} & Cs & 0.972(2) & 0.979(3)\\
        & Rb & 0.987(6) & 0.994(7)\\
        \multirowcell{2}{Ground state detection $D_\mathrm{g}$\\($t_\mathrm{drop} = \SI{1.5}{\micro s}$)} & Cs & 0.959(3) & 0.986(5)\\
        & Rb & 0.966(5) & 0.978(8)\\
        \multirowcell{2}{Rydberg state detection $D_\mathrm{r}$} & Cs & 0.99(1) & 1.00(1)\\
        & Rb & 0.94(1) & 0.94(1)\\
    \end{tabular}
    \caption{Overview of SPAM calibration parameters, as extracted directly from uncorrected measurement data (raw) and after correcting for other SPAM errors (corrected) as indicated in the respective paragraph.}
    \label{tab:spam_parameters}
\end{table}

\subsubsection{Correction procedure}

We now discuss how experimentally-obtained data is corrected for the preparation and readout errors introduced above.

In our experiments, we measure binary outcomes ${x_i\in\left\{0,1\right\}}$ on $N$ atoms, with $i=1,\dots,N$. This results in $2^N$ potential outcomes $\bm{x}=(x_1,\dots,x_N)\in\left\{0,1\right\}^N$. From $n_\text{tot}$ measurements, wherein string $\bm{x}$ was observed $n_{\bm{x}}$ times, we extract the occurrence probabilities $p_{\bm{x}}=n_{\bm{x}}/n_\text{tot}$, thereby obtaining a probability vector $\bm{p}$ (such that $1\geq p_{\bm{x}}\geq 0$ and $\sum_{\bm{x}}p_{\bm{x}}=1$). This is affected by SPAM errors.

Our goal is to estimate the underlying \emph{true} probability distribution $\bm{P}$, if there were no SPAM errors at play. This is related to the measured distribution through a linear error map,
\begin{equation}
    \bm{p} = M\bm{P},
\end{equation}
where $M$ is the SPAM error matrix, constructed as follows.

First, it is useful to introduce the following measurement error probabilities $e_0$ and $e_1$, defined respectively as
\begin{align}
    e_0 & = D_gF_n + (1-D_g)(1-F_p),  \nonumber \\
    e_1 & = D_rF_p + (1-D_r)(1-F_n); &
\end{align}
these correspond, respectively, to the probability of erroneously assigning the measurement outcomes $1$ or $0$. Then, for a single atom the SPAM error map can be written as 
\begin{equation}
\begin{aligned}
    A = F\eta &\left[\begin{matrix} 1-e_0 & e_1 \\
        e_0 & 1-e_1
    \end{matrix}\right] \\ 
    + F(1-\eta) &\left[\begin{matrix} 1-e_0 & 1-e_0 \\
        e_0 & e_0
    \end{matrix}\right] \\
    + (1-F) &\left[\begin{matrix} e_1 & e_1 \\
        1-e_1 & 1-e_1
    \end{matrix}\right],
\end{aligned}
\end{equation}
where we dropped the labels indicating the species. Thus, we define one such map per species, $A_\mathrm{Cs}$ and $A_\mathrm{Rb}$.

For many atoms, the simplest SPAM error map reads 
\begin{equation}
    M = \dots \otimes A_\mathrm{Cs} \otimes A_\mathrm{Rb} \otimes A_\mathrm{Cs} \otimes A_\mathrm{Rb} \otimes \dots,
\end{equation}
wherein we take the tensor product of the single atom maps. This treats all the atoms independently, in fact neglecting correlated errors between individual traps. We find that, given our error parameters, the leading contributions of SPAM errors are indeed well captured by the model above, as correlated effects impact our results only beyond our current experimental precision. This is also certified through extensive numerical simulations, which show how the form above for $M$ is able to faithfully capture our SPAM errors.

 From the map above, it is then straightforward to numerically extract an estimate of the true probability distribution $\bm{P}$ from the experimentally measured $\bm{p}$. Importantly, since $M$ is local, we are able to implement such SPAM correction even for relatively large system sizes, whenever the observed quantities are also local (e.g. for the cluster state stabilizers).

\vspace{2cm} 
\bibliography{QCA}